\documentclass{aa}
\usepackage{psfig}
\begin{document}
\title{Controlling Artificial Viscosity in SPH simulations of accretion disks}
\author{Annabel Cartwright and Dimitrios Stamatellos}
\institute{            School of Physics and Astronomy,Cardiff University, 5, The Parade, Cardiff CF24 3AA}

\date{}

\abstract
{The fidelity of Smoothed Particle Hydrodynamics (SPH) simulations of accretion disks depends on how Artificial Viscosity (AV) is formulated.}{We investigate whether standard methodology is reliable in this regard.}{We test the operation of two methods for selective application of AV in SPH simulations of Keplerian Accretion Disks, using a ring spreading test to quantify effective viscosity, and a correlation coefficient technique to measure the formation of unwanted prograde alignments of particles.} {Neither the Balsara Switch ($B$) nor Time Dependent Viscosity (TDV) work effectively, as they leave AV active in areas of smooth shearing flow, and do not eliminate the accumulation of alignments of particles in the prograde direction. The effect of both switches is periodic, the periodicity dependent on radius and unaffected by the density of particles. We demonstrate that a very simple algorithm activates AV only when truly convergent flow is detected and reduces the unwanted formation of prograde alignments. The new switch works by testing whether all the neighbours of a particle are in Keplerian orbit around the same point, rather than calculating the divergence of the velocity field, which is very strongly affected by Poisson noise in the positions of the SPH particles.}

\keywords{Viscosity, Accretion disks, Angular momentum}
\authorrunning{A. Cartwright}
\titlerunning{A switch for Viscosity in SPH simulations}
\maketitle


\section{Introduction}

In Smoothed Particle Hydrodynamics (SPH) simulations which involve strongly convergent flows, it is essential to include Artificial Viscosity (AV) (Monaghan $\&$ Gingold 1983; Monaghan $\&$ Lattanzio 1985; Monaghan 1992). It causes close, approaching SPH particles to repel each other, with a force which increases with approach velocity. The result is that particles in colliding streams are rapidly decelerated and the streams do not pass through each other. Such particle interpenetration would be an unphysical situation and cannot be allowed in a simulation. It is important to note that AV was never intended to replicate the behaviour of real viscosity (Monaghan 2005).

SPH is now used widely to simulate the evolution of disks (Bate et al 2003; Lodato $\&$ Rice 2004; Rice et al 2003a, 2003b, 2004; Mayer et al 2004; Schaefer et al 2004; Boffin et al 1998; Watkins et al 1998a and 1998b, Stamatellos et al 2008 and 2009). The behaviour of viscosity in real circular shear flow in fluids is understood analytically and experimentally (eg Feynman 1964), and the evolution of accretion disks is predicted to be dominated by effects such as turbulence which behave like viscosity, causing angular momentum to be transferred outwards, enabling matter to fall in and accrete onto a central object (Shakura $\&$ Sunyaev 1973; Lynden Bell $\&$ Pringle 1974). It is often assumed that turbulence is generated and moderated by the magneto-rotational instability (Balbus $\&$ Hawley 2002).

However, AV, if allowed to operate unchecked in a simulated Keplerian Disk, does not operate only when convergence occurs, but also acts to slow down neighbouring particles as they overtake because of the differential rotation. The magnitude of AV is therefore, systematically, too high. Clarke (2009) has shown that it is a key requirement for correct modelling of the evolution of disks that numerical viscosity is kept very low, otherwise it dominates the pseudo viscous forces, such as self-gravity, in the simulation. It is therefore crucial to disable AV in normal, equilibrium Keplerian shear flow, and to activate it only when it is needed to capture a shock. 

In this paper we set out to test the Balsara Switch ($B$), and Time Dependent Viscosity (TDV), for their effectiveness in disabling AV in non-convergent Keplerian shear flow. In Sect. 2 we explain the theoretical operation of $B$ and in Sect. 3 we examine the actual values of $B$ calculated for a snapshot of a smoothly rotating accretion disk. In Sect. 4 we investigate the variation of $B$ with time for individual SPH particles.
In Sect. 5 we turn to TDV and look at actual values calculated for a simulated Keplerian disk in which there is no convergence. In Sect. 6 we look at the spreading of a ring of particles, as a quantitative measure of the effectiveness of $B$ and TDV in reducing viscous shear. We also quantify, for the first time, the previously reported tendency for AV to cause SPH particles to form prograde alignments. A new approach to controlling AV in disks, by simply checking whether neighbours are in Keplerian orbit around the same object, is explained and evaluated in Sect. 7 and these results are discussed in Sect. 8.

\section{The Balsara Switch}

The Balsara Switch ($B$) is a well known tool for enabling the rapid application of AV in the presence of a shock (Balsara 1989), and works as follows. A multiplicative `Balsara factor',

\begin{equation} \label{balsara}
 B_i= \frac{\mid \nabla \cdot {\bf v}\mid_i}{\mid\nabla\times {\bf v}\mid_i+\mid\nabla \cdot  {\bf v} \mid_i}
\end{equation}
is calculated for each SPH particle $i$, where  $\bf v$ is the velocity field and the derivatives are evaluated at position ${\bf r}_i$. 

The viscous acceleration $\Pi_{ij}$ between each pair of particles is then multiplied by the mean of the values of $B$, thus
\begin{equation}
{\Pi}_{ij}\to\frac{(B_i + B_j)}{2} { \Pi}_{ij}
\end{equation}

In a Keplerian disk, with all matter orbiting in centrifugal balance about a mass $M$, ${\mid\nabla\times \bf v\mid}$ and ${\mid\nabla \cdot \bf v\mid}$ at radius $r$ are given by :

\begin{eqnarray} \label{kepvel2} \nonumber
{\mid\nabla\times \bf v\mid}&=  &\left(\frac{GM}{4r^3}\right)^{1/2} \\ 
{\mid\nabla \cdot \bf v\mid}&= &0.
\end{eqnarray}
${\mid\nabla\times \bf v\mid}$ is therefore finite while ${\mid\nabla \cdot \bf v\mid}$ is zero, and hence ${B}$ is zero for all $r$. However, if particles start to converge at a particular location, ${\mid\nabla \cdot \bf v\mid}$ increases, so ${B}$ increases, approaching a limiting value of 1.0, when ${\mid\nabla \cdot \bf v\mid \gg \mid\nabla\times \bf v\mid}$. Multiplying the viscosity coefficients by $B$ should effectively switch off the AV when a disk is rotating stably with a Keplerian velocity profile, but switch it on, selectively, for any particles in an area of true convergence.

\section{Noise problems with the SPH implementation of $B$}

In SPH simulations $B$ is calculated for each particle using Eqn.~\ref{balsara}, substituting numerical estimates $\nabla \cdot {\bf v \mid_{SPH}}$ and $\nabla \times {\bf v \mid_{SPH}}$ for the true values of $\nabla \cdot \bf v$ and $\nabla \times \bf v$ :

\begin{eqnarray} \label{divv} 
 \nabla \cdot {\bf v}_i \mid_{\bf SPH}&= &\sum_j \frac{m_j}{\rho_j} 
\frac{{\bf v}_{ij}\cdot{\bf r}_{ij}}{\mid{\bf r}_{ij}\mid h_{ij}^4}
{W'\left(\frac{\mid{\bf r}_{ij}\mid}{h_{ij}}\right)},\\
 \nabla \times {\bf v}_i \mid_{\bf SPH}&= &\sum_j \frac{m_j}{\rho_j} \frac{{\bf v}_{ij}\times{\bf r}_{ij}}{\mid{\bf r}_{ij}\mid h_{ij}^4}{W'\left(\frac{\mid{\bf r}_{ij}\mid}{h_{ij}}\right)},
\end{eqnarray} 

\begin{figure} 
\centerline{\psfig{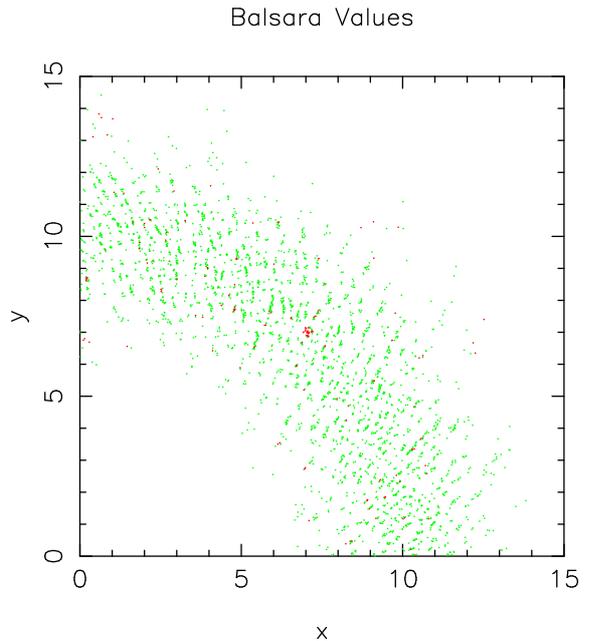}}

\caption{Values of $B$ for a portion of a ring of particles in Keplerian orbits. }
\label{bals1}
\end{figure}

From Eqn.~\ref{kepvel2}, the value of $B$ should be zero for a stably rotating disk of particles
in Keplerian orbits, where there is no convergence. However, as Fig. ~\ref{bals1} demonstrates, this does not happen in practice in SPH simulations. 

Figure~\ref{bals1} shows that high values of $B$ were calculated for 80,000 SPH particles in purely 
Keplerian orbits around a star mass $M_* =1M_{\odot}$. The two dimensional ring was initialised with  randomly positioned particles, such that the surface density was Gaussian around a maximum value at radius = 10AU. An area of convergence was created around coordinates (7,7), by adding convergent velocities up to 1.5km/s to the Keplerian velocities, which are around 150km/s at these radii.  In Fig. ~\ref{bals1} a quadrant of the ring was viewed face-on, the particles orbiting in the anti-clockwise direction. Particles with $B>0.3$ were plotted as green dots, and those with $B>0.6$ as red dots. Clearly $B$ was found to be high, not only in the converging region, but all over the disk, with some values as high as 0.8.  The particles with high $B$ were not scattered randomly over the disk, rather they appeared to form alignments, predominantly in the `leading' direction. 

The neighbours of a particle in a Keplerian disk fall into four quadrants of approaching or receding particles. Particles ahead and in outer orbits and particles behind and on inner orbits are approaching, so ${\bf v}_{ij}\cdot{\bf r}_{ij}$ is negative. Particles ahead on inner orbits and behind on outer orbits are receding, so ${\bf v}_{ij}\cdot{\bf r}_{ij}$ is positive. The calculation of $\nabla \cdot \bf v\mid_{SPH}$ adds together positive values of ${\bf v}_{ij}.{\bf r}_{ij}$ from receding neighbours and negative values from  approaching ones. If neighbours are evenly distributed, $\nabla \cdot \bf v\mid_{SPH}$ is small, as the positive and negative components cancel out. However if neighbours are unevenly distributed, Poisson noise causes $\nabla \cdot \bf v\mid_{SPH}$ to increase in magnitude.

This is shown in column 2 of Table ~\ref{bvsvg}, where the mean value of $B$ was calculated for two disks. The first disk contained randomly positioned particles in perfect Keplerian orbits, and $B$ should therefore be zero. An average value of $B=0.11$ was consistently found in this and all other randomly populated disks, for numbers of particles ranging from 80000 to 1 million. The second, settled, disk was created by allowing the first to evolve using pressure and AV, controlled by $B$. There was no convergence in this disk, and because the particles were more evenly spaced, $B$ was closer to zero. $B$ fell to less than half the value of the original disk, to $B=0.052$, and remained steady. Column 3 of the same table shows the mean value of $B$ calculated for a convergence zone created within the same settled disk.

\begin{table}
\caption{\footnotesize Mean value of $B$ and $\cal{NK}$ for 80,000 SPH particles orbiting in a Keplerian disk.}
\label{bvsvg}
\vspace{0.5cm}
\normalsize
\begin {center}
\begin {tabular} {lcccc} 
Disk&$B$&$B$& ${\cal NK}$&${\cal NK}$\\
Status&Disk &Conv&Disk&Conv\\
\hline
& &  &  & \\
Random&0.11 	&   	&0.003	& \\
Settled&0.052	&0.116	&0.060	&0.774\\
\end {tabular}
\end {center}
\end{table}


\section{Variation of $\nabla \cdot \bf v \mid_{SPH}$ with time}

In a Keplerian disk, all particles are constantly overtaking their outer neighbours and being overtaken by inner neighbours. As a result, three or more particles often find themselves in radial alignment. In this case,  these aligned neighbours are in opposite quadrants, so they are all receding or all approaching.  They therefore contribute to a large negative value of $\nabla \cdot \bf v\mid_{SPH}$ while they are approaching radial alignment, and then a high positive value, once they have passed radial alignment and are spreading out. This explains the orientation of the linear features observed in Fig.~\ref{bals1}. Chance alignments of particles will produce high values of $B$ for any orientation of alignments, if $B$ is calculated for both positive and negative values of $\nabla \cdot \bf v\mid_{SPH}$. However if $B$ is set to zero for positive values of $\nabla \cdot \bf v\mid_{SPH}$, high $B$ values only occur for points in alignments with outer particles ahead and inner particles behind.

\begin{figure}
\centerline{\psfig{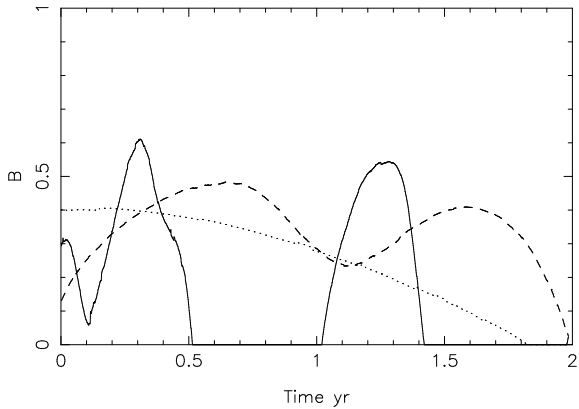}}

\caption{$B$ plotted against time, for SPH particles at three different radii in a Keplerian disk around a $1M_{\odot}$ star.  }
 \label{bvst}
\end{figure}

Figure~\ref{bvst} shows the value of $B$ for three SPH particles in different orbits within a Keplerian disk. As $B$ has been set to zero when $ \nabla \cdot \bf v\mid_{SPH}$ is positive, we see the positive half of three periodical functions.The solid line indicates a particle at radius 1.4AU, the dashed line a particle at 3 AU and the dotted line a particle at 10AU. Orbital periods for these radii are 1.6y, 5y and 30y. The discontinuities in $B$ are due to modifications to the neighbour lists, which affect the calculated values of $\nabla \cdot \bf v\mid_{SPH}$ and $\nabla \times \bf v\mid_{SPH}$. Clearly $B$ is varying quasi-periodically, with a frequency decreasing with the radius of the orbit.

This is explained by considering the `overtaking' speed of a neighbour. Consider a particle orbiting mass $M$ at radius $r$, and its neighbour orbiting at radius $r+h$.  The overtaking orbital speed, $v_{rel}$, of the inner particle is 

\begin{equation} \label{overtake} 
v_{rel}\simeq \left\vert \frac{dv_{orb}}{dr}\right \vert h = \frac{\left(GM\right)^{1/2}}{2r^{3/2}}h
\end{equation}

In order to overtake the neighbour, we allow the particle to travel a distance $h$ further along its orbit than its neighbour does, which takes a time $\frac{h}{v_{rel}}$. During this time the distance travelled is therefore 

\begin{equation} \label{overtake2}
\frac{h}{v_{rel}} \cdot v_{orb} = 2r
\end{equation}
which is approximately one third of an orbit. 

If the inner particle travels $2r$ along its orbit, the outer must travel $2r + 2h$, just to keep its station. So, adding this effect to that calculated in Eqn. \ref{overtake2}, we have the particle moving $3h$ past a neighbour a distance $h$ away in the time it takes to travel one third of the way round its orbit. This should be approximately the period of one cycle of the value of $B$.

The variation of  $\nabla \cdot \bf v \mid_{SPH}$, and hence of $B$, with time is therefore predicted to be periodic with frequency approximately three times the orbital frequency. A particle orbiting at radius 1.4 AU with an orbital period of 1.6 year should show 4 cycles of $B$ in a 2 year period. For 3 AU the period is about 5 years and we expect just over one cycle. For 10 AU, with an orbital period 30 years, we only expect to see about $\frac{1}{5}$ of the cycle. This agrees well with what is seen in Fig. \ref{bvst}.

Once an aligned structure has formed, all the particles in that structure will have high  $\nabla \cdot \bf v \mid_{SPH}$, values, and therefore both $B$ and the magnitude of the AV (which is also calculated using  $\nabla \cdot \bf v \mid_{SPH}$), will be large for all these aligned particles.  They will therefore tend to stick to their neighbours more than other particles. The longer such structures exist, the more angular momentum they will transfer. Imaeda and Inutsuka (2002) observed that linear structures are a pronounced feature of SPH simulations of shear flow which use AV. This raises the worrying possibility that AV itself could be enhancing the formation of filaments in disk simulations, with $B$ doing nothing to moderate this tendency. This could lead to misleading results.

\section{Time Dependent Viscosity}
A second method for selectively switching on AV in shocks while keeping it switched off at other times, was proposed by Morris and Monaghan (1997), and is referred to as Time Dependent Viscosity (TDV).

Each SPH particle $i$ is given its own viscosity parameter $\alpha_i$, which evolves with time, according to the equation 

\begin{equation} \label{tdf}
{\frac{d\alpha_i}{dt}= - \frac{\alpha_i-\alpha^*}{\tau_i}+ S_i},
\end{equation}
where $S_i$ is the source term,

\begin{equation} \label{source}
{S_i= max(-\nabla \cdot {\bf v\mid_{SPH}},0)}.
\end{equation}
$\tau_i$ is the e-folding time, which is set to be

\begin{equation} \label{tau}
{\tau_i=\frac{h_i}{C_1 c_i}},
\end{equation}
where $h_i$ is the smoothing length for viscous and pressure forces, $c_i$ the sound speed and $C_1$ a parameter chosen to control the rate at which $\alpha_i$ decays back to $\alpha^*$.

Clearly $\alpha_i$ increases if $\nabla \cdot \bf v \mid_{SPH}$ is negative, and decays to a minimum value $\alpha^*$ otherwise. This provides a rapidly  increasing viscosity around a particle which is entering a shock, where $\nabla \cdot \bf v\mid_{SPH}$ is large and negative for a number of successive timesteps. Once the shock is passed, $\nabla \cdot \bf v\mid_{SPH}$ falls to small or even positive values, and the viscosity dies away again.  Morris and Monaghan (1997) suggest that setting $C_1$ to $0.2$ should ensure that the viscosity dies down to the minimum value within about 5 smoothing lengths.

This method is not helpful for preventing spuriously high viscosity in a Keplerian accretion disk, the problem being that once again $\nabla \cdot \bf v\mid_{SPH}$ is being used as an indicator of convergence. As we have shown, $\nabla \cdot \bf v\mid_{SPH}$ varies periodically with a frequency dependent on radius, as particles overtake one another, causing alignments to form. It is impossible to tune the TDV method to filter out such low frequency noise without making it much too slow to react to shocks.

\section{AV Tests}
\subsection{Spreading ring test}

As a simple test of the magnitude of effective shear viscosity in a real SPH simulation, with and without $B$ or TDV, a ring spreading test was carried out. Lynden-Bell and Pringle (1974) predicted that a ring of particles will spread under the influence of viscosity, the rate of spreading increasing with viscosity. Murray (1996) and Flebbe et al.(1994) conducted SPH simulations to demonstrate this effect, starting with a ring initialised with a Gaussian Surface Density profile. 

The SEREN SPH code (Hubber 2010, in prep) was used. This is a code specifically developed for the study of star formation. A ring of 80,000 SPH particles, total mass 0.1$M_{\odot}$ was created, with a Gaussian distribution of surface density about a maximum at radius 10 AU. This was set in Keplerian orbit about a 1$M_{\odot}$ star, with no self gravity within the ring and no pressure forces, and allowed to evolve only under the influence of the star's gravity and AV, for a fixed simulated real time duration (200 years).  Multiple Particle Time Stepping was used with a maximum range of $10^4$ from the shortest to the longest timesteps. As the ring evolved, all the particles were binned in 100 bins by radius, the mean surface density calculated for each bin, and the maximum value stored as $\Sigma_{max}(t)$. AV was calculated for 50 neighbours, and only applied for negative values of $\nabla \cdot \bf v\mid_{\rm SPH}$. The viscosity parameter $\alpha$ was varied from 0.1 to 1.0, while $\beta$ was kept at 0.0.  This was then repeated with $B$ implemented, as defined in Eqn.~\ref{balsara} and again with TDV. The results are shown in Table ~\ref{efold}. Column 1 gives the value of AV $\alpha$ parameter, column 2 the method used to moderate AV, ( $B$, TDV, ${\cal NK}$ or NONE). Column 3 gives the measured reduction in maximum surface density, $\delta\Sigma(\%)$.

\begin {table}  
\caption{\footnotesize Reduction in maximum surface density, $\delta\Sigma(\%)$ found for various AV prescriptions in the ring spreading test. }
\label{efold}
\normalsize
\begin {center}
\begin {tabular}  [tbp]{lcc} 
$\alpha$& AV Switch&$\delta\Sigma(\%)$\\
 \hline
0.0 	&None 	&0.00	\\
0.1 	&None  	&1.08	\\
0.5 	&None 	&2.18	\\
1.0 	&None 	&3.39	\\

\hline
1.0 	&BS 	&0.72	\\
1.0 	&TDV 	&0.92	\\
1.0 	&${\cal NK}$	&0.00	\\

\hline

\end {tabular}
\end {center}

\end{table}

The rings did not evolve smoothly, the central density oscillating as the ring evolved. However, there was a measurable trend, and the first four rows of Table~\ref{efold} show that the ring spread more quickly, with the central density falling more rapidly, with increasing values of $\alpha$. 

With no convergence within the ring, $d\Sigma$ should be $0.0$ if $B$ and TDV were effective in switching off AV in shear flow. Table~\ref{efold} shows that implementing $B$ and keeping $\alpha=1.0$ did decrease the effective viscosity, giving a smaller $\delta\Sigma$ than that obtained when $\alpha$ was reduced to 0.1. By comparison, our mean $B$ value for 50 neighbours in Table~\ref{bvsvg} was 0.052. TDV worked slightly less well, $\delta \Sigma$ being larger than for $B$.

\begin{figure}
\centerline{\psfig{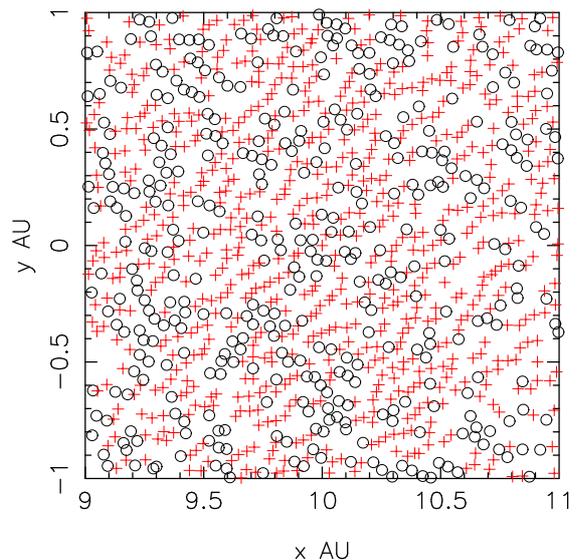}}

\caption{ Prograde (red crosses) and retrograde (black circles) particle alignment in a disk evolved with pressure and viscous forces.  }

\label{alignf}
\end{figure}
\subsection{Measurement of particle alignment}
We have suggested that prograde alignments of particles will persist because they are held together by AV, which is artificially high in these alignments. The data from the ring spreading experiments was therefore analysed to quantify any such effect. Each particle in turn was examined, together with any close neighbours, defined as lying within 0.5$h$, where $h$ is the SPH smoothing length. The coordinates of the close neighbours were projected into radial and tangential components, within the disk, and if at least three points were found within $0.5h$ the correlation coefficient $cc$ was calculated for the group of points. A value of $cc>0.5$ indicated a linear alignment in the prograde direction, while $cc<-0.5$ indicated a linear alignment in the retrograde direction. Figure ~\ref{alignf} illustrates that this calculation correctly identified the prograde and retrograde alignments (red crosses and black circles respectively) in a portion of the disk, rotating anticlockwise. 

Prograde and retrograde alignments were then counted, and the results are shown in column 7 of table ~\ref{align}. In the disk illustrated in Fig.~\ref{alignf}, 24$\%$ of particles were in prograde alignments, and 10$\%$ in retrograde, a prograde surplus of 14$\%$. 

\begin {table}  
\caption{\footnotesize Percentages of prograde ($\delta A^+$) and retrograde ($\delta A^-$) alignments found in disks evolved with various prescriptions of AV.}
\label{align}
\normalsize
\begin {center}
\begin {tabular}  [tbp]{lccccc} 
$\alpha$& Pressure& AV Switch&$\delta A^+(\%)$&$\delta A^-(\%)$\\
 \hline
0.0 	&No	&None 	&18	&18\\
0.1 	&No	&None  	&26	&14\\
0.5 	&No	&None 	&28	&12\\
1.0 	&No	&None 	&28	&11\\

\hline
1.0 	&No	&$B$ 	&27	&14\\
1.0 	&No	&TDV 	&28	&13\\
1.0 	&No	&${\cal NK}$	&19	&18\\

 \hline
0.0 	&Yes	&None 	&20	&20\\
0.1 	&Yes	&None  	&24	&15\\
0.5 	&Yes	&None 	&22	&15\\
1.0 	&Yes	&None 	&35	&11\\

\hline
1.0 	&Yes	&$B$ 	&30	&12\\
1.0 	&Yes	&TDV 	&32	&12\\
1.0 	&Yes	&${\cal NK}$	&23	&17\\
\hline

\end {tabular}
\end {center}

\end{table}
Initial, random, particle arrangements had $18\%$ particles in prograde and $18\%$ in retrograde alignments, so there was no excess in prograde alignments in the initial disk. The first four rows of Table~\ref{align} show that there were also no excess prograde alignments when the disk was evolved with no AV ($\alpha=0.0$), but when AV was applied, the number of prograde alignments increased to $26-28\%$ while the number of retrograde alignments decreased to $11-14 \%$. The next three rows show that using $B$ resulted in $27\%$/$14\%$ prograde excess, and the TDV AV $28\% /13\%$. Thus, although both methods moderated the spreading effect of AV, the build up of an excess of prograde alignments was still large.

The trial was repeated with the addition of pressure forces. This increased the number and imbalance of alignments. With $\alpha=0.0$, no effective AV, the number of alignments was seen to increase in both directions to $20\%$. As AV was increased the number of prograde alignments increased and retrograde decreased, the largest imbalance being seen for $\alpha=1.0$, when $35\%$ of particles were in prograde alignments, and only $11\%$ in retrograde. $B$ and TDV produced only slightly fewer alignments than this. Thus pressure forces were seen to exacerbate the problem of prograde alignments being formed preferentially while retrograde alignments either did not form or were disrupted.
 
\section{Keplerian recognition approach} \label{patrec}
Another way of ruling out areas of apparent convergence which are in fact simply regions of Keplerian shear, is to compare neighbours' positions and velocities with a Keplerian template. Here we use the facts that in a Keplerian disk, the velocity is always perpendicular to the position vector of a particle relative to the centre of rotation, and the quantity $v^2r$ is constant over all of the disk. Clearly, the position of the centre of rotation must be known to apply these tests.

For each particle $p_i$, we first calculated $v_i^2r_i$ for the particle itself. We then examined each of the ${\cal{N}}_{NEIB}$  neighbours of the particle. A neighbour was counted as anomalous if its velocity was not perpendicular to its position vector, indicated by $ {\bf r_j}.{\bf v_j}/ |r_jv_j| > \epsilon$. We also counted the point as anomalous if ${v^2_i r_i /v^2_j r_j}>1.0+\epsilon$ or $<1.0-\epsilon$. The error value $\epsilon$ yielded good discrimination when set to a value of $0.03$. Large $\epsilon$ failed to identify important discrepancies from Keplerian flow, while very small $\epsilon$  caused viscosity to be switched on when pressure forces caused slightly non-Keplerian velocity gradient.

The number of anomalous particles ${\cal{N}}_A$ were accumulated and then divided by the total number of neighbours to give a viscosity switch $\cal{NK} ={\cal{N}}_A/{\cal{N}}_{NEIB}$. $\cal{NK}$ was therefore near zero for regions where all particles were in Keplerian orbit around the same point, and increased to 1 as more and more neighbours fail to conform to a Keplerian pattern. Table ~\ref{bvsvg} shows that in a disk of 80000 particles orbiting with a perfectly Keplerian velocity profile, the mean value of $\cal{NK}$ was 0.003. When the disk was allowed to settle, using both pressure and AV, $\cal{NK}$ increased to 0.06. Although this mean value is slightly greater than the mean value of $B$,  the distribution of high values was very different. High values of $B$ were registered all over the disk, while high values of $NK$ were mainly found in the edge regions.  

\begin{figure}
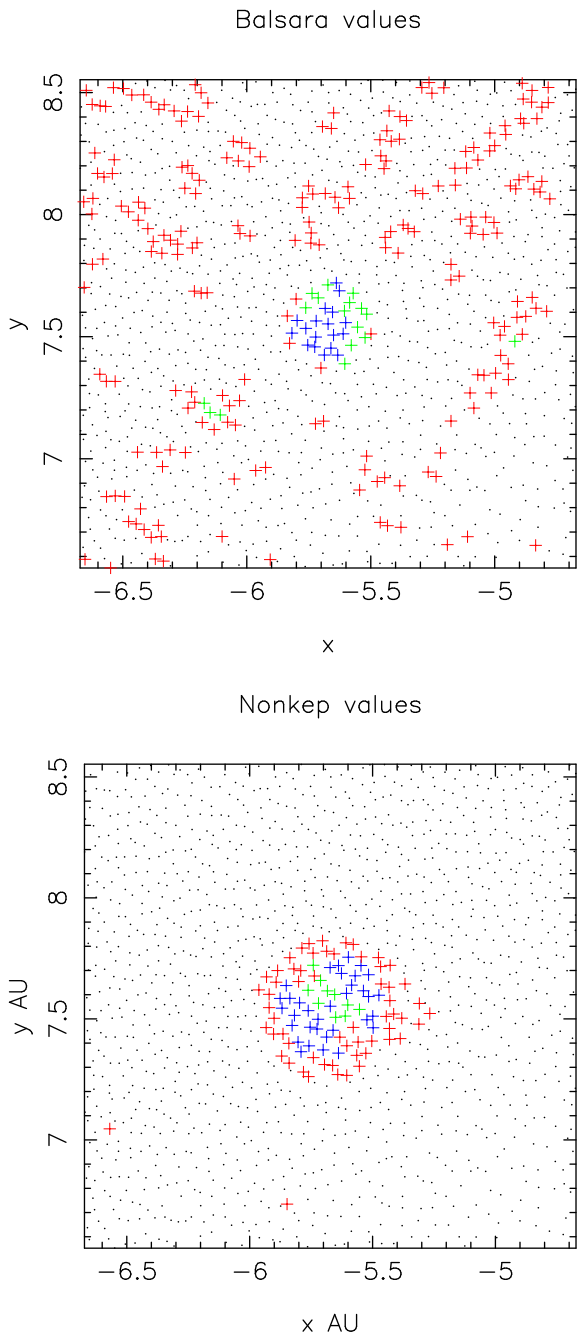

\centerline{\psfig{figure=fig4a.ps,width=7.5cm,angle=270}}

\vspace{.5cm}
\centerline {\psfig{figure=fig4b.ps,width=7.5cm,angle=270}}

\caption{$B$ (top) and $\cal{NK}$ (bottom) values for points in a Keplerian disk, with an area of converging flow superimposed at coordinates (-5.6,7.6).  }
 \label{patt}
\end{figure}

$\cal{NK}$ was found to be effective at activating AV in areas of convergence. This is illustrated in Fig. ~\ref{patt}, where $\cal{NK}$ is compared with $B$. A ring was evolved with particles in Keplerian orbits with pressure and AV activated, and then  a circular region of converging velocities was added to the Keplerian flow. All fifty neighbours of a point were given a convergence velocity proportional to their distance from the central point and pointing towards it,  with the maximum convergence velocity being $1\%$ of the orbital velocity of the central particle. This was added to the orbital velocity of each particle. Note that this was quite a stiff test, the magnitudes of the artificial converging velocities being of the same order of magnitude as those due to the Keplerian velocity shear within the convergence zone, and much smaller than the sound speed. The particles in the region of the convergence are shown, with crosses to indicate particles for which $B$ or the non-Keplerian indicator $\cal{NK}$, were greater then 0.2, green indicates values higher than 0.5, blue greater than 0.8. It can be seen that both methods correctly indicate the area of convergence, but $\cal{NK}$ increases for all particles in the convergence region. $B$ is raised for most particles in the convergence zone, but is also incorrectly raised for alignments of particles which are not in the convergence zone. The mean value of $\cal{NK}$ within the convergence zone is 0.78, and .06 outside while mean $B$ is only 0.116 inside the zone and 0.052 outside. 

The ring spreading test was repeated using the $\cal{NK}$ factor, and the results are presented in Table ~\ref{efold}. $\cal{NK}$ kept AV switched off and the ring did not spread at all. There was also only a very small increase in the growth of prograde linear alignments (from $18$ to $19\%$) when the $\cal{NK}$ factor was used. When pressure forces were added to the simulation, the $\cal{NK}$ factor did not completely suppress the formation of alignments, but performed much better than $B$ and TDV. 

\begin{figure*}
\centerline{\psfig{figure=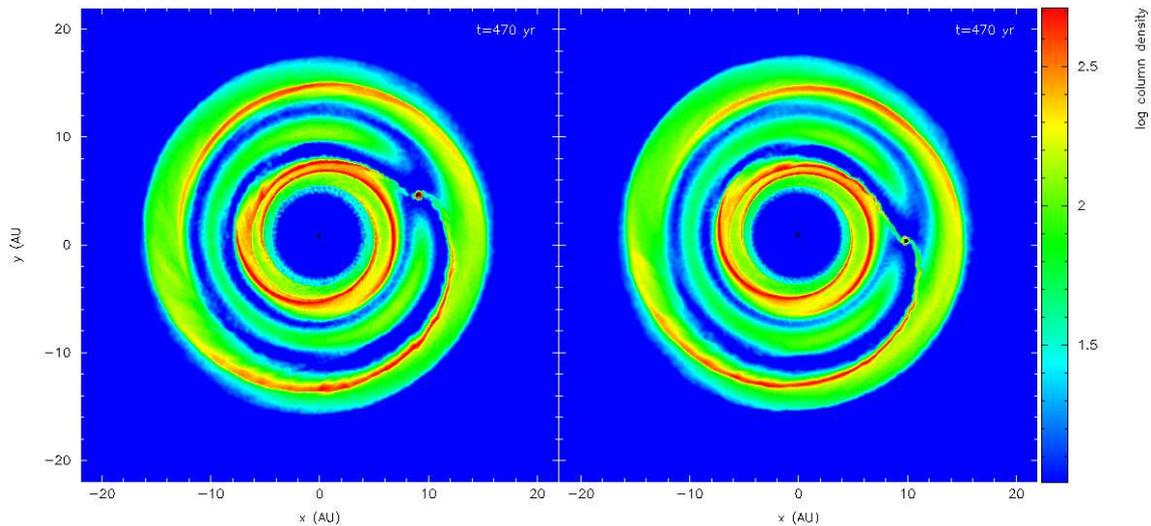,width=15cm,angle=270}}

\caption{Disks incorporating a planetary mass object, evolved using left, Balsara Switch and right the $\cal{NK}$ factor.}

 \label{planet}
\end{figure*}

Finally, to test the $\cal{NK}$ method with a more realistic simulation, an isothermal disk was evolved containing a Jupiter-mass object.  Initial conditions were isothermal, T=10K, surface density $\Sigma=\Sigma(1 AU) (R/au)^{-7/4}$, $M_{disk}=0.01 M_{\odot}$. $M_{star}=1M_{\odot}$. The disk extended from 5 to 15 au and planet mass was 1 $M_{J}$ at radius 10 au. 500,000 SPH particles were used and the simulation ran for 470yrs, about 15 orbital periods of the planet. The results are shown in Fig.~\ref{planet}. It can be seen that $\cal{NK}$ was effective at capturing the shocks, and the predicted spiral wave was observed. For comparison, the same simulation was performed using $B$, and the results are similar, but the spirals and shocks are slightly better defined when using the $\cal{NK}$ method.

\section{Discussion}
The SPH estimate of the divergence of viscosity ${\mid\nabla \cdot \bf v\mid}$, (Eqn.~\ref{divv}) when used in a Keplerian Disk, is dominated by the Poisson noise arising from the imperfectly spaced SPH particles (Cartwright et al 2009). As the estimate of ${\mid\nabla \cdot \bf v\mid}$ is the key factor in calculating both $B$ and TDV, both of these methods erroneously register the apparent convergence of overtaking neighbours as a signal that AV should be applied. 

$B$ is prone to an alignment artifact, varying periodically with a frequency $\sim $ 3 times the orbital frequency, which is the overtaking frequency for neighbours. This is independent of $h$ and therefore independent of the number of SPH particles used. 

TDV is also not ideal for use in Keplerian disks if used with $\nabla \cdot \bf v\mid_{SPH}$, as a source term. $\nabla \cdot \bf v\mid_{SPH}$ is the source of the low frequency alignment artifact which compromises $B$, and applying a low pass filter to it does not remove the effect. 

AV if applied, even at quite a low level, in a Keplerian Disk, results in the formation of alignments of particles in the prograde direction. This can be quantified by calculating the correlation coefficient $cc$ of close neighbours of particles, positive values $cc>0,5$  indicating prograde alignments, $cc<-0.5$ indicating retrograde alignments. A randomly populated disk of particles has no excess of either, having about $18\%$ of each. The balance is unaffected if the disk is evolved with only pressure forces and no AV. However, as soon as AV is used, even at low levels using $B$, the particles start to form an excess of $12-16\%$ prograde alignments. Using pressure forces in addition to AV causes an even larger surplus of alignments to appear, up to $24\%$ being measured when maximum AV was implemented. This should be of great concern for the disk-modelling community. Normally, disk simulations result in the formation of spiral arms in the retrograde direction, which may then form gravitational instabilities. The action of AV to form and preserve prograde alignments, while destroying retrograde ones, may well be interfering in this process.

In order to prevent particle inter-penetration, AV must be applied promptly when areas of convergence are detected. $B$ is not as effective here as might be expected. When an area of convergence is superimposed upon Keplerian shear, the mean $B$ value within the shear zone is only $B=0.116$, whereas values greater than this are seen elsewhere in the undisturbed regions of the disk.

A solution to the problem of keeping AV switched off in regions of Keplerian Shear, and reaching the very low levels pointed out as necessary by Clarke (2009), may lie in a pattern recognition approach. Here we demonstrate a very simple method, which checks that all neighbours of a point are travelling orthogonally to the radius of the disk, and that the value of $v^2r$ is constant within a small tolerance for all neighbours. This simple method results in AV being switched off consistently in areas of smooth Keplerian shear (mean ${\cal{NK}}=0.06$), while switching on more effectively in areas  of simulated convergence (${\cal{NK}}=0.77$ in convergence zone). It also markedly reduces the formation of prograde alignments of SPH particles. In a disk evolution simulation with pressure and AV enabled, $B$ and TDV resulted in $22\%$ and $24\%$ surplus prograde alignments, while the excess with the $\cal{NK}$ switch was only $6\%$. The $\cal{NK}$ method requires the orientation and centre of rotation of the disk to be known, but this is true for most disk simulations, and easily deduced for others. The calculation is no more complicated than that for the $B$, so there is no increased computational overhead, and it is very amenable to parallelisation. ${\cal{NK}}$ also has the interesting feature of being variable, by altering the value of $\epsilon$. A high value of $\epsilon\simeq 0.1$ will keep AV switched off in normal thermal motions superimposed on Keplerian shear, but will only switch on for particle $i$ if neighbouring particles $j$ have non-Keplerian approach velocities $v_{ij}$ greater than $10\%$ of $v_i$. If $\epsilon \simeq 0.01$, AV will be applied to smaller discrepancies from Keplerian motion, which may include some thermal motions within the disk.

Note that $\cal{NK}$ does not identify convergence, merely the absence of Keplerian flow. However, as it is then used in conjunction with AV, which does detect convergence, the two methods in combination are effective in applying a decelerating viscous force only when SPH particles are in truly convergent flow within a Keplerian disk.

Finally, it should be noted that the flaw in the calculation of ${\mid\nabla \cdot \bf v\mid}$ also applies in the calculation of ${\mid\nabla \cdot \bf a\mid}$. Large divergence in the acceleration or velocity fields can apparently be detected when in fact the variation is due to noise in the particle positioning. Both quantities should be used with caution, particularly as diagnostic indicators in the creation of sink particles.

\section{Conclusion}
Both $B$ (Balsara 1989) and the TDV approach (Morris and Monaghan 1997) to controlling AV in circular shear flow fail for the same reasons: SPH estimates of $\nabla \cdot {\bf v}$ appear to detect convergence in steady shear flow. AV controlled by these methods is therefore applied in all areas of a simulated Keplerian accretion disk, with results which will be unprepresentative of real viscous evolution, and compromise the results (Clarke, 2009). AV results in the selective formation of prograde alignments of SPH particles, which can be measured using a correlation coefficient method. We find that a disk populated with randomly placed SPH particles initially has 18$\%$ particles in prograde alignments and 18$\%$ retrograde. After being allowed to evolve using pressure forces and AV, these figures change to 28$\%$ and 12$\%$. Clearly the use of AV changes the arrangement of particles in the experiment, even when very low levels of AV are applied. 

An alternative method of controlling AV, based on identifying Keplerian velocity characteristics, shows promise, being more effective both at switching on in regions of convergence and off in pure Keplerian flow, and significantly reducing the accumulation of excess prograde alignments of particles.

\section{Acknowledgements}Annabel Cartwright is the holder of a Royal Society Dorothy Hodgkin Fellowship. Some of this work was undertaken as part of a PPARC funded studentship. D.Stamatellos acknowledges post-doctoral support from STFC under the auspices of the Cardiff Astronomy Rolling Grant. Grateful thanks to Anthony Whitworth for his guidance and insight.


\end{document}